\documentclass[12pt]{iopart}
\usepackage{iopams}
\expandafter\let\csname equation*\endcsname\relax
\expandafter\let\csname endequation*\endcsname\relax
\usepackage{array,amsmath}
\pdfoutput=1
\usepackage{graphicx,psfrag,color}
\usepackage[utf8]{inputenc}
\usepackage[T1]{fontenc}
\usepackage{lmodern}
\usepackage{times}
\usepackage{mathptmx}
\usepackage{epstopdf}
\usepackage{xcolor}
\usepackage[export]{adjustbox}
\usepackage[colorlinks,linkcolor=blue,urlcolor=blue,citecolor=blue]{hyperref}
\usepackage[normalem]{ulem}
\usepackage{xspace}
\usepackage{float}
\usepackage{bm}
\usepackage{cite}
\usepackage{lineno}
\hypersetup{
    colorlinks=true,
    linkcolor=blue,
    filecolor=magenta,
    urlcolor=cyan,
    citecolor=cyan
}

\DeclareSymbolFont{bbold}{U}{bbold}{m}{n}
\DeclareSymbolFontAlphabet{\mathbbold}{bbold}
\makeatletter
\newcommand*\bigcdot{\mathpalette\bigcdot@{.55}}
\newcommand*\bigcdot@[2]{\mathbin{\vcenter{\hbox{\scalebox{#2}{$\m@th#1\bullet$}}}}}
\makeatother

\def\ee{\mathrm{e}}

\begin{document}
\newcommand{\gae}{\lower 2pt \hbox{$\, \buildrel {\scriptstyle >}\over {\scriptstyle
\sim}\,$}}
\newcommand{\lae}{\lower 2pt \hbox{$\, \buildrel {\scriptstyle <}\over {\scriptstyle
\sim}\,$}}

\title{Coherence-decoherence interplay in quantum systems due to projective stochastic pulses: The case of Rabi oscillations}
\author{Sushanta Dattagupta$^1$ and Shamik Gupta$^2$}
\address{$^1$Sister Nivedita University, Kolkata, West Bengal 700156, India\\$^2$Department of Theoretical Physics, Tata Institute of Fundamental Research, Homi Bhabha Road, Mumbai 400005, India}
\ead{sushantad@gmail.com,shamik.gupta@theory.tifr.res.in}
\date{\today}

\begin{abstract}  
The interplay of coherence and decoherence is played out in a three-level quantum system, in which the third level is incoherently coupled to the second one which itself is in coherent interaction with the first level. The study is based on a stochastic scenario in which the coherent, unitary evolution of the system is randomly interrupted by a Poisson-driven pulse sequence. In the absence of an external pulse, the system undergoes coherent, unitary evolution restricted to the subspace spanned by the first level (level $1$) and the second level (level $2$). The application of a pulse induces transitions between the second and the third level (level $3$), thereby introducing non-unitary effects that perturb the otherwise isolated two-level dynamics. The pulses are assumed to have infinitesimal duration, with strengths modeled as random variables that are uncorrelated across different pulses. A representative model for the stochastically-averaged transition (super)operator mimicking the dynamics induced by the application of pulses allows for an analytical derivation of the matrix elements of the averaged density operator, especially, the element pertaining to the first level. When the system is initially in level $1$, we obtain in particular the temporal behavior of the stay-put probability, that is, the probability $P_1(t)$ that the system is still in level $1$ at time $t$. As a function of time, the quantity $P_1(t)$ exhibits a coherence-to-decoherence crossover behavior. At short times $t \ll 1/\lambda$, where $\lambda$ is the average frequency at which pulses are applied to the system, coherent dynamics dominate. Consequently, $P_1(t)$ displays pronounced Rabi-like oscillations. At long times $t \gg 1/\lambda$, decoherence effects prevail, leading to an exponential decay of the form $P_1(t)-1/3 \sim \exp(-\lambda t)$. The results are relevant in the context of well-established Rabi oscillations, applicable to molecular beams and magnetic resonance experiments, as well as in more contemporary frameworks including quantum information processing, quantum resetting protocols, and the theory of quantum trajectories.
\end{abstract}
\date{\today}
\maketitle
\tableofcontents

\section{Introduction}
\label{sec:setup}

Isolated quantum systems evolve coherently according to the Schr\"{o}dinger equation~\cite{Streltsov:2017}. A central focus in contemporary quantum research concerns study of open quantum systems~\cite{Breuer:2002} and in particular the issue of how coupling a quantum system to an external environment affects quantum coherence and introduces the undesirable effects of decoherence~\cite{Zurek:2003,Schlosshauer:2005}. This question is especially relevant in the context of quantum technology~\cite{Pellizzari:1995}, and has gained further importance thanks to recent experimental advances in atomic, molecular, and optical systems that allow for controlled investigation of nonequilibrium dynamics of open quantum many-body systems~\cite{Bloch:2008,Syassen:2008,Barreiro:2011,Ritsch:2013,Daley:2014}.

There are two primary approaches to studying the dynamics of an open quantum system, in the Schr\"{o}dinger picture. The more conventional method involves formulating an effective description for the time evolution of the density operator of the system via a quantum master equation, typically in the form of the so-called Lindblad equation~\cite{Breuer:2002,Manzano:2020,Ding:2024}. This approach relies on several uncontrolled approximations, such as assuming weak coupling between the system and the environment, as well as a clear separation of time scales between the intrinsic dynamics of the system and that of the environment. The second approach is motivated by standard experimental protocols, wherein the evolution of a quantum system alternates between intervals of effective isolation and intervals during which environmental interactions affect its dynamics. A common example is shot noise, which randomly disrupts the unitary evolution of the system~\cite{Dattagupta:2013,Nikolaenko:2023}. This type of dynamics is modeled as unitary evolution interrupted at random times by non-unitary evolution due to interactions with the environment. As a result, the system undergoes a sequence of unitary evolutions that are stochastically modulated over time~\cite{Clauser:1971,Dattagupta:1977,Christensen:1979,Dattagupta:1987,Dattagupta:1991,Das:2022,Dattagupta:2022,Gzyl:2024}. In this framework, no assumptions are required regarding the strength of the coupling to the environment, and the approximation of time-scale separation is naturally embedded in the dynamics. This work focuses on the second approach and the insights it provides into the fate of quantum coherence under the influence of environmental interactions.

To proceed, let us consider a very general set-up comprising a quantum system described by a given time-dependent Hamiltonian $H(t)$. The system is at zero temperature. Working in units in which the Planck's constant is unity, we consider the following dynamical scenario: the system is undergoing unitary evolution in time, which is interspersed at random instants of time distributed according to a Poisson distribution with application of pulses of infinitesimal duration, up to a certain fixed time $t$. Our set-up follows the one proposed in Ref.~\cite{Clauser:1971}. The evolution of the system between two successive instants of the application of pulses is unitary and hence coherent, while non-unitarity and hence decoherence is induced repeatedly in the evolution at time instants of the application of pulses. 
Thus, starting with a density operator $\rho(0)$, a unitary
evolution for a random time is followed by an instantaneous interaction with the pulses and which is modelled in terms of a time-independent interaction (super)operator $\mathcal{T}$.  The result undergoes unitary evolution for another random time, followed by another instantaneous interaction, and so on.  Then, over the time interval $[\,0,\,t\,]$, a realization of the  evolution involves $p \ge 1$ number of application of pulses at
random time instants $t_1,t_2,\ldots,t_p$, with the time duration $\tau_{p'+1} \equiv t_{p'+1} -t_{p'};~p'=0,1,2,\ldots,p-1;~\,t_0=0$ between
successive interactions being random variables sampled independently from an exponential distribution $p(\tau)$:
\begin{align}
p(\tau)=\lambda\ee^{-\lambda\tau},
\label{eq:ptau-exponential}
\end{align}
where $\lambda>0$ is the inverse of the average time between two
successive applications of pulses. In other words, $\lambda$ is the average frequency of application of pulses. The evolution ends with unitary evolution for time duration $t-t_p$. 

We take the time-dependent Hamiltonian of the system as
\begin{align}
    H(t)=H_0+\sum_{p\ge 1}V_p\delta(t-t_p),
\label{eq:Ht}
\end{align}
where the time-independent part $H_0$ dictates  the unitary evolution of the system, while the time-dependent part $\sum_{p\ge 1}V_p\delta(t-t_p)$, arising due to the application of pulses, appears only in the interaction operator $\mathcal{T}$. The average density
operator (averaged over different realizations of the dynamics outlined above) at time $t$ reads as~\cite{Das:2022}
\begin{align}
\overline{\rho}(t) &= \sum_{p=0}^\infty \int_0^t {\rm d}t_p \int_0^{t_p} {\rm d}t_{p-1}\ldots \int_0^{t_3}{\rm d}t_2 \int_0^{t_2} {\rm d}t_1 \, F(t-t_p) \, e^{-i H_0^\times (t-t_p)} \nonumber \\
&\times \mathcal{T}p(t_p-t_{p-1})e^{-i H_0^\times (t_p-t_{p-1})}\mathcal{T}\ldots
\mathcal{T}p(t_2-t_1)e^{-i H_0^\times (t_2-t_1)}\mathcal{T} p(t_1)e^{-i H_0^\times t_1}\rho(0) 
\label{eq:rho-evolution-10} \\
&= U(t)\rho(0),
\label{eq:rho-evolution-11}
\end{align}
where $H_0^\times$ is the time-independent Liouvillian (super)operator corresponding to the Hamiltonian $H_0$, and $F(t)=\int_t^\infty d\tau~p(\tau)$ is the probability
for no interaction to occur during time $t$. Here, we have defined $U(t)$ as the time-evolution (super)operator. Note that we are working in the Schr\"{o}dinger picture in which states and density operators evolve in time, as opposed to the Heisenberg approach in which states and density operators are time independent, while operators standing for dynamical variables are time dependent and follow evolution according to the so-called quantum Langevin equation~\cite{Ghosh:2024}. Considering Laplace transform of both sides of Eq.~\eqref{eq:rho-evolution-11}, one obtains in the Laplace space the equation~\cite{Das:2022}
\begin{align}
\widetilde{\overline{\rho}}(s)= \widetilde{U}(s)\rho(0), \label{eq:rho-in-lap}
\end{align}
with $s$ being the Laplace variable,  
\begin{align}
 \widetilde{U}(s)=\left[(s+\lambda)\mathbb{I}+i H_0^\times-\lambda\mathcal{T}\right]^{-1},
        \label{eq:Us}
\end{align}
and $\mathbb{I}$ being the identity operator.

As a specific example of a dynamics described by Eq.~\eqref{eq:Us}, which has the structure emerging in a stochastic Liouville equation set-up~\cite{Dattagupta:1987,Kubo:1969}, let us take the static Hamiltonian $H_0$ to represent a two-state problem -- in particular a spin-half system -- in the presence of a magnetic field of strength $B_0$ along the $z$-axis and a magnetic field of strength $B$ along the transverse $x$-axis. Thus, we have
\begin{align}
H_0 = - B_0 \sigma_z - B \sigma_x, 
\label{eq:Rabi-1}
\end{align}
where the $\sigma$'s are the Pauli matrices. Denoting the eigenstates of $\sigma_z$ by $|1\rangle$ and $|2\rangle$ and assuming that the spin is prepared in state $|1\rangle$ at time $t = 0$, so that the initial density matrix is $\rho(0)=|1\rangle\langle 1|$, the probability of finding the spin in state $|2\rangle$ at time $t$ is~\cite{Sakurai:1965}
\begin{align}
P_2(t)=\langle 2|\rho(t)|2\rangle=\frac{B^2}{B^2+B_0^2}\sin^2\left(t \sqrt{B^2 + B_0^2}\right).
\label{eq:Rabi}
\end{align}
The oscillatory behavior encapsulated in the above equation, which is directly linked with quantum coherence, is characteristic of the Rabi formula that is a generic feature of molecular beams, masers, and so forth~\cite{Sakurai:1965}. The central aim of our present study is to investigate how, given that initially the system was in state $|1\rangle$, the stay-put probability $P_1(t)(= 1 - P_2(t))$ is affected because of decoherence induced by a neighboring state through which a stochastic drive is triggered. In other words, in the set-up of Eq.~\eqref{eq:Ht}, the system is subject to pulses at Poisson-distributed random times that induce transitions between the upper state $|2\rangle$ and the neighboring state.  The presence of the neighboring state, the third level $|3\rangle$, replenishing the upper level is reminiscent of how a laser works. Normally, in a laser, the upper level undergoes a ‘population inversion’ because of stimulation provided by a third level. In our scenario, however, we are dealing with a quantum system at zero temperature, wherein population inversion has no relevance. Furthermore, unlike the typical case in lasers where the third level coherently pumps the upper level, in our setup, the third level acts as a source of quantum dissipation. Specifically, coupling to the third level via applied pulses induces non-unitary evolution in the system that would otherwise evolve unitarily.  

Let us state our main results at the outset. In this work, we derive in the Laplace domain an exact analytical expression for the stay-put probability $P_1(t)$, Eq.~\eqref{eq:P1s}. This quantity exhibits a remarkable crossover behavior: As a function of time, the quantity exhibits a clear crossover between two distinct dynamical regimes. At short times $t \ll 1/\lambda$, coherent dynamics dominate, and $P_1(t)$ displays pronounced Rabi-like oscillations. At longer times, decoherence effects prevail, leading to an exponential decay of the form $P_1(t) \sim \exp(-\lambda t)$, see Fig.~\ref{fig:fig2}. Here, as discussed above, $\lambda$ is the inverse of the average time between two subsequent application of pulses. 

The paper is organized as follows. In Section~\ref{sec:model}, we describe our model in detail in Subsection~\ref{sec:model-subsec} and proceed towards analyzing it using the stochastic Liouville equation approach described above. Here, our objective is to compute the stay-put probability $P_1(t)$, which we take up in Subsection~\ref{sec:stayput}. The subsection is further divided into several small subsections, Section~\ref{sec:T} -- Section~\ref{sec:U0s}, each of which is devoted to computation of various terms that contribute to the final expression for $P_1(t)$ in the Laplace domain, Eq.~\eqref{eq:P1s}. We then check in Subsection~\ref{sec:check} certain known limiting cases of our derived result. The following subsection is devoted to analyzing and discussing the behavior of the stay-put probability and highlighting in particular the coherence-to-decoherence crossover behavior that it exhibits. The paper ends with conclusions in Section~\ref{sec:conclusion}. 

\section{Our model system and its analysis}
\label{sec:model}

\subsection{The model}
\label{sec:model-subsec}

We now come to the specifics of our model system. Consider a three-level closed quantum system at zero temperature, with energy levels $|1\rangle$, $|2\rangle$, and $|3\rangle$, and with $|1\rangle \langle 1|+|2\rangle \langle 2|+|3\rangle \langle 3|=\mathbb{I}$, the $3 \times 3$ identity operator. Also, the states $|1\rangle$, $|2\rangle$, and $|3\rangle$ form a complete set of orthornormal states.  The system under its unitary evolution can access only levels $1$ and $2$, and not level $3$. However, at random instants of time distributed according to a Poisson distribution, the system is subject to pulses of infinitesimal duration that induce transitions between levels $2$ and $3$. In other words, if the system prior to the application of a pulse is in level $2$, the pulse would project the system to level $3$, and vice versa. In the framework of Eq.~\eqref{eq:Ht}, we  have 
\begin{align}
 H_0 \equiv \epsilon_1|1\rangle \langle 1|+\epsilon_2|2\rangle \langle 2|+\epsilon_3|3\rangle \langle 3|+\Delta (|1\rangle \langle 2|+|2\rangle \langle 1|) 
 \label{eq:H0}
\end{align}
as the bare Hamiltonian of the system that leads to its coherent evolution between two time instants of the application of pulses, with $\epsilon_1$, $\epsilon_2$ and $\epsilon_3$ being the energies of the three levels. The term proportional to $\Delta$ results in transitions between levels $1$ and $2$. Measuring the energies $\epsilon_1$ and $\epsilon_2$ with respect to $\epsilon_3$, we have
\begin{align}
 H_0=\epsilon_1|1\rangle \langle 1|+\epsilon_2|2\rangle \langle 2|+\Delta (|1\rangle \langle 2|+|2\rangle \langle 1|). 
 \label{eq:H}
\end{align}

On the other hand, the operators $V_p$ are given by
\begin{align}
V_p=\theta_p(|2\rangle \langle 3|+|3\rangle \langle 2|)=\theta_p S;~S \equiv |2\rangle \langle 3|+|3\rangle \langle 2|.
\label{eq:Vp}
\end{align}
Here, $\theta_p$, a random variable, denotes the strength of the pulse that is applied at time instant $t_p$. We take the random variables $\theta_p$ to be completely uncorrelated for different $p$'s. The random times $t_p$ and the random pulse strengths $\theta_p$ are the two sources of (classical) randomness in the dynamics, besides the quantum randomness arising due to inherent quantum fluctuations. 

Starting at time $t=0$ with the system in level $1$, our question of interest is: what is the state of the system at any time $t>0$? More specifically, what is the stay-put probability $P_1(t)$ of the system in level $1$ at time $t>0$? Evidently, from the very nature of the studied dynamical scenario, one expects an interplay of time scales related to coherent evolution of the system and to decoherence effects induced due to the application of pulses. Depending on which one of the two time scales is dominant would dictate the state of the system at a later time $t>0$, and it is this feature of interplay that we want to unravel in our present study.

\subsection{Analysis of the stay-put probability}
\label{sec:stayput}

Considering the form of Eq.~\eqref{eq:Vp}, we see that the interaction operator $\mathcal{T}$ for our model system is itself stochastic, in that it contains the classical stochastic variable $\theta$ that takes up a different value every time there is an application of pulse. Recall also that this stochastic variable (hence, the interaction operator $\mathcal{T}$) corresponding to different instants of application of pulses is completely uncorrelated. Then, while computing the average density operator, one has to  average further the result~\eqref{eq:rho-evolution-10} with respect to the statistical properties of this source of randomness. Denoting the corresponding averaged interaction operator by $(\mathcal{T})_\mathrm{av}$, and following the same steps as the ones involved in deriving Eq.~\eqref{eq:Us}, we obtain in the Laplace space the time-evolution operator, in which all sources of randomness have been averaged over, as
\begin{align}
 \widetilde{U}(s)=\left[(s+\lambda)\mathbb{I}+i H_0^\times-\lambda(\mathcal{T})_\mathrm{av}\right]^{-1};~~\mathcal{T}=e^{-i\theta S^\times},
        \label{eq:Us-0}
\end{align}
where $S^\times$ is the time-independent Liouvillian (super)operator corresponding to the operator $S$ in Eq.~\eqref{eq:Vp}. Reverting to the time domain gives
\begin{align}
    \overline{\rho}(t)=U(t)\rho(0);~U(t)=\exp[(-iH_0^\times+\lambda(\mathcal{T})_\mathrm{av}-\lambda)t].
    \label{eq:rhot-evaluate}
\end{align}
The difference of the present work with respect to Ref.~\cite{Das:2022} is that in the latter, the only source of randomness was in the time instants of interaction, and there was no further randomness as contained in our case in the stochastic variable $\theta$.

\subsubsection{Matrix elements of superoperator $\mathcal{T}$}
\label{sec:T}

We begin our analysis with the computation of the matrix elements of the superoperator $\mathcal{T}$. As a model for the random variable $\theta$, we take it from dimensional arguments to be an angle-like variable distributed uniformly in $[0,2\pi)$, yielding $(\cos \theta)_\mathrm{av}=(\sin \theta)_\mathrm{av}=0$, and $(\cos^2 \theta)_\mathrm{av}=(\sin^2 \theta)_\mathrm{av}=1/2$. Noting that 
$S^2=|2\rangle \langle 2|+|3\rangle \langle 3|=P_2+P_3$, the respective projection operators, and $S^3=S,S^4=S^2,S^5=S$, etc., we get
\begin{align}
    e^{-i \theta S}=\mathbb{I}-(P_2+P_3)(1-\cos \theta)-i S \sin \theta.
\end{align}
Using the basis states as $|m\rangle;~m=1,2,3$ and the rules associated with computing the matrix elements of a superoperator~\cite{Dattagupta:1987,Das:2022}, we have
\begin{align}
    (m_1 m_2|\mathcal{T}|m_3 m_4)&=\langle m_1|e^{-i \theta S}|m_3\rangle \langle m_4|e^{i \theta S}|m_2\rangle \nonumber \\
    &=\left(\delta_{m_1 m_3}-\langle m_1|(P_2+P_3)|m_3\rangle (1-\cos \theta)-i \sin \theta \langle m_1|S|m_3\rangle \right)\nonumber \\
    &\times \left(\delta_{m_2 m_4}-\langle m_4|(P_2+P_3)|m_2\rangle (1-\cos \theta)+i \sin \theta \langle m_4|S|m_2\rangle \right).
\end{align}
Using the above equation and performing the average over $\theta$, it follows straightforwardly on using the orthonormality of the set of states $\{|m\rangle;~m=1,2,3\}$ that
\begin{align}
    (\mathcal{T})_\mathrm{av}=\mathcal{T}_1+\Delta \mathcal{T},
    \label{eq:T-av}
\end{align}
where the $9 \times 9$ matrix $\mathcal{T}_1$ has its only nonzero element given by $(11|\mathcal{T}_1|11)=1$, i.e., 
\begin{align}
    (m_1m_2|\mathcal{T}_1|m_3 m_4)=\delta_{m_1 m_3}\delta_{m_2 m_4}\delta_{m_1 m_2}\delta_{m_1 1},
\end{align}
while the $9 \times 9$ matrix $\Delta \mathcal{T}$ has just eight non-zero elements, each equal to $1/2$, given by
\begin{align}
&(22|\Delta \mathcal{T}|22) = (33|\Delta \mathcal{T}|33) = (22|\Delta \mathcal{T}|33) = (33|\Delta \mathcal{T}|22) \nonumber \\
&=(23|\Delta \mathcal{T}|23) = (32|\Delta \mathcal{T}|32) =(23|\Delta \mathcal{T}|32) = (32|\Delta \mathcal{T}|23) =1/2. \nonumber \\
\label{eq:DeltaT}
\end{align}

\subsubsection{Matrix elements of superoperator $H_0^\times$}
\label{sec:H0x}

We now come to the computation of the matrix elements of the superoperator $H_0^\times$. Using Eq.~\eqref{eq:H}, we have \begin{align}
    H_0=\begin{bmatrix} \epsilon_1& \Delta & 0 \\ \Delta & \epsilon_2 & 0 \\ 0 &0& 0 \end{bmatrix},
    \label{eq:H0-matrix-1}
\end{align}
with the matrix labeled by the states $|1\rangle$, $|2\rangle$ and $|3\rangle$. This yields
\begin{align}
    (m_1m_2|H_0^\times|m_3m_4)=\langle m_1|H_0|m_3\rangle \delta_{m_2m_4}-\langle m_4|H_0|m_2\rangle \delta_{m_1m_3},
\end{align}
so that we have 
\begin{align}
    H_0^\times=\begin{array}{c@{}}
(11| \\ (22| \\ (12| \\ (21| \\ (33| \\ (13| \\ (23| \\ (31| \\ (32|
\end{array}
&\begin{bmatrix} 
    0& 0 & -\Delta & \Delta & ~~~0 & ~~~~0 & ~~~0 & ~~~0 & ~~~0 \\
    0& 0 & \Delta & -\Delta & ~~~0 & ~~~~0 & ~~~0 & ~~~0 & ~~~0 \\
    -\Delta & \Delta & \epsilon_{12} & 0 & ~~~0 & ~~~~0 & ~~~0 & ~~~0 & ~~~0 \\
    \Delta & -\Delta & 0 & -\epsilon_{12} & ~~~0 & ~~~~0 & ~~~0 & ~~~0 & ~~~0 \\
    0& 0 & 0 & 0 & ~~~0 & ~~~~0 & ~~~0 & ~~~0 & ~~~0 \\
    0& 0 & 0 & 0 & ~~~0 & ~~~~\epsilon_{1} & ~~~\Delta & ~~~0 & ~~~0 \\
    0& 0 & 0 & 0 & ~~~0 & ~~~~\Delta & ~~~\epsilon_{2} & ~~~0 & ~~~0 \\
    0& 0 & 0 & 0 & ~~~0 & ~~~~0 & ~~~0 & ~~~-\epsilon_{1} & ~~~-\Delta \\
    0& 0 & 0 & 0 & ~~~0 & ~~~~0 & ~~~0 & ~~~-\Delta & ~~~-\epsilon_{2}
    \end{bmatrix},\nonumber \\
        &\begin{array}{@{}*{9}{wc{7mm}}@{}}
~~~~~~|11) & ~~~~~~|22) & ~~~~~~|12) & ~~~~~~|21) & ~~~~~~|33) & ~~~~~~~~|13) & ~~~~~~~~|23) & ~~~~~~~~~~|31) & ~~~~~~~~~~~~~~~~|32)\\ 
\label{eq:H0-matrix-2}
\end{array} 
\end{align}
which shows that the $9 \times 9$ matrix for $H_0^\times$ breaks up into two $4 \times 4$ block matrices with non-zero elements. Here, we have defined $\epsilon_{12} \equiv \epsilon_1-\epsilon_2$.

On the other hand, with $\mathbb{I}$ being the $9 \times 9$ identity matrix, we have $(\mathcal{T})_\mathrm{av}-\mathbb{I}=\mathcal{T}_1+\Delta \mathcal{T}-\mathbb{I}$, which is a $9 \times 9$ matrix given below, see the discussion in Section~\ref{sec:T}:
\begin{align}
    (\mathcal{T})_\mathrm{av}-\mathbb{I}=\begin{array}{c@{}}
(11| \\ (22| \\ (12| \\ (21| \\ (33| \\ (13| \\ (23| \\ (31| \\ (32|
\end{array}
&\begin{bmatrix} 
    0& 0 & 0 & ~~~~0 & ~~~0 & ~~~~0 & ~~~0 & ~~~0 & ~~~0 \\
    0& -1/2 & 0 & ~~~~0 & ~~~1/2 & ~~~~0 & ~~~0 & ~~~0 & ~~~0 \\
    0 & 0 & -1 & ~~~~0 & ~~~0 & ~~~~0 & ~~~0 & ~~~0 & ~~~0 \\
    0 & 0 & 0 & ~~~~-1 & ~~~0 & ~~~~0 & ~~~0 & ~~~0 & ~~~0 \\
    0& 1/2 & 0 & ~~~~0 & ~~~-1/2 & ~~~~0 & ~~~0 & ~~~0 & ~~~0 \\
    0& 0 & 0 & ~~~~0 & ~~~0 & ~~~~-1 & ~~~0 & ~~~0 & ~~~0 \\
    0& 0 & 0 & ~~~~0 & ~~~0 & ~~~~0 & ~~~-1/2 & ~~~0 & ~~~1/2 \\
    0& 0 & 0 & ~~~~0 & ~~~0 & ~~~~0 & ~~~0 & ~~~-1 & ~~~0 \\
    0& 0 & 0 & ~~~~0 & ~~~0 & ~~~~0 & ~~~1/2 & ~~~0 & ~~~-1/2
    \end{bmatrix}.\nonumber \\
        &\begin{array}{@{}*{9}{wc{7mm}}@{}}
~~~~~~|11) & ~~~~|22) & ~~|12) & ~~~~~~|21) & ~~~~~~~~~~~~~~|33) & ~~~~~~~~~~~~~~~~~~~~|13) & ~~~~~~~~~~~~~~~~~~~~~~~~~~~~|23) & ~~~~~~~~~~~~~~~~~~~~~~~~~~~~~~~~~~~~~~|31) & ~~~~~~~~~~~~~~~~~~~~~~~~~~~~~~~~~~~~~~~~~~~~~~~~|32)\\ 
\label{eq:H0-matrix-22}
\end{array} 
\end{align}
One may check that the eigenvalues of the above matrix are $-1$ with six-fold degeneracy and $0$ with three-fold degeneracy. Using the fact that the matrix $(-iH_0^\times+\lambda(\mathcal{T})_\mathrm{av}-\lambda)$ has its largest eigenvalue equal to zero, with the corresponding unnormalised eigenvector being $(1,1,0, 0, 1, 0, 0, 0, 0)^\mathrm{T}$, with T denoting transpose, we conclude on using Eq.~\eqref{eq:rhot-evaluate} that 
\begin{align}
    \overline{\rho}(t\to \infty)=\begin{bmatrix}
    1/3&0&0\\
    0&1/3&0\\
    0&0&1/3
    \end{bmatrix}.
    \label{eq:rhoasymp}
\end{align}

\subsubsection{Matrix elements of superoperator $\widetilde{U}(s)$}
\label{sec:Us}
Using Eq.~\eqref{eq:T-av} in Eq.~\eqref{eq:Us-0}, we get 
\begin{align}
 \widetilde{U}(s)=\left[(s+\lambda)\mathbb{I}+i H_0^\times-\lambda \mathcal{T}_1-\lambda \Delta \mathcal{T}\right]^{-1},
\label{eq:Us-12}
\end{align}
which on expanding in powers of $\lambda$ yields the Dyson series:
\begin{align}
    \widetilde{U}(s)=\widetilde{U}_0(s)\left[\mathbb{I}+\lambda \Delta \mathcal{T}\widetilde{U}_0(s)+\lambda^2\Delta \mathcal{T}\widetilde{U}_0(s)\Delta \mathcal{T}\widetilde{U}_0(s)+\ldots\right],
    \label{eq:series-1}
\end{align}
where we have 
\begin{align}
 \widetilde{U}_0(s) \equiv \left[(s+\lambda)\mathbb{I}+i H_0^\times-\lambda \mathcal{T}_1\right]^{-1}.
        \label{eq:Us-3}
\end{align}

Our quantity of interest, namely, the stay-put probability for level $1$, is given by 
\begin{align}
    P_1(t)=\langle 1|\overline{\rho}(t)|1\rangle,
\end{align}
which in the Laplace space reads as
\begin{align}
    \widetilde{P}_1(s)=\langle 1|\widetilde{U}(s)\rho(0)|1\rangle,
\end{align}
on using Eq.~\eqref{eq:rho-in-lap}. Note that we have 
\begin{align}
    \rho(0)=|1\rangle \langle 1|.
    \label{eq:IC}
\end{align}
We thus need to compute the quantity
\begin{align}
    \langle 1|\widetilde{U}(s)\rho(0)|1\rangle&=\sum_{m_1 m_2}(11|\widetilde{U}(s)|m_1m_2)\langle m_1|\rho(0)|m_2\rangle\nonumber \\
    &=(11|\widetilde{U}(s)|11)\nonumber \\
    &=(11|\widetilde{U}_0(s)\left[\mathbb{I}+\lambda \Delta \mathcal{T}\widetilde{U}_0(s)+\lambda^2\Delta \mathcal{T}\widetilde{U}_0(s)\Delta \mathcal{T}\widetilde{U}_0(s)+\ldots\right]|11),
    \label{eq:Us-1}
\end{align}
where we have used Eq.~\eqref{eq:series-1}.

With $s'=s+\lambda$, we have $(s+\lambda)\mathbb{I}+iH_0^\times-\lambda \mathcal{T}_1$
\begin{align}
=\begin{array}{c@{}}
(11| \\ (22| \\ (12| \\ (21| \\ (33| \\ (13| \\ (23| \\ (31| \\ (32|
\end{array}
&\begin{bmatrix} 
    s& 0 & -i\Delta & i\Delta & ~~~0 & ~~~~0 & ~~~0 & ~~~0 & ~~~0 \\
    0& s' & i\Delta & -i\Delta & ~~~0 & ~~~~0 & ~~~0 & ~~~0 & ~~~0 \\
    -i\Delta & i\Delta & s'+i\epsilon_{12} & 0 & ~~~0 & ~~~~0 & ~~~0 & ~~~0 & ~~~0 \\
    i\Delta & -i\Delta & 0 & s'-i\epsilon_{12} & ~~~0 & ~~~~0 & ~~~0 & ~~~0 & ~~~0 \\
    0& 0 & 0 & 0 & ~~~s' & ~~~~0 & ~~~0 & ~~~0 & ~~~0 \\
    0& 0 & 0 & 0 & ~~~0 & ~~~~s'+i\epsilon_{1} & ~~~i\Delta & ~~~0 & ~~~0 \\
    0& 0 & 0 & 0 & ~~~0 & ~~~~i\Delta & ~~~s'+i\epsilon_{2} & ~~~0 & ~~~0 \\
    0& 0 & 0 & 0 & ~~~0 & ~~~~0 & ~~~0 & ~~~s'-i\epsilon_{1} & ~~~-i\Delta \\
    0& 0 & 0 & 0 & ~~~0 & ~~~~0 & ~~~0 & ~~~-i\Delta & ~~~s'-i\epsilon_{2}
    \end{bmatrix}.\nonumber \\
        &\begin{array}{@{}*{9}{wc{7mm}}@{}}
~~~~~~|11) & ~~~~~~|22) & ~~~~~~~~~~~~~~|12) & ~~~~~~~~~~~~~~~~~~~~~~~~~~|21) & ~~~~~~~~~~~~~~~~~~~~~~~~~~~~~~~~|33) & ~~~~~~~~~~~~~~~~~~~~~~~~~~~~~~~~~~~~~~~~~~~~|13) & ~~~~~~~~~~~~~~~~~~~~~~~~~~~~~~~~~~~~~~~~~~~~~~~~~~~~~~~~~~~~~|23) & ~~~~~~~~~~~~~~~~~~~~~~~~~~~~~~~~~~~~~~~~~~~~~~~~~~~~~~~~~~~~~~~~~~~~~~~~|31) & ~~~~~~~~~~~~~~~~~~~~~~~~~~~~~~~~~~~~~~~~~~~~~~~~~~~~~~~~~~~~~~~~~~~~~~~~~~~~~~~~~~~~~~~~~|32)\\ 
\label{eq:H0-matrix-2}
\end{array} 
\end{align}
It then follows due to the block diagonal structure of $(s+\lambda)\mathbb{I}+iH_0^\times-\lambda \mathcal{T}_1$ that for the matrix $\widetilde{U}_0(s)=\left[(s+\lambda)\mathbb{I}+i H_0^\times-\lambda \mathcal{T}_1\right]^{-1}$, we have $(11|\widetilde{U}_0(s)|mn)\ne 0$ only when either $m=n=1$ or $m=n=2$ or $m=1,n=2$ or $m=2,n=1$. In particular, marking the non-zero elements of $\widetilde{U}_0(s)$ with a star, we have  
\begin{align}
\widetilde{U}_0(s)=\begin{array}{c@{}}
(11| \\ (22| \\ (12| \\ (21| \\ (33| \\ (13| \\ (23| \\ (31| \\ (32|
\end{array}
&\begin{bmatrix} 
    *& ~~* & ~~~~* & ~~~~* & ~~~~~0 & ~~~~~~0 & ~~~~~0 & ~~~~~0 & ~~~~~0 \\
    *& ~~* & ~~~~* & ~~~~* & ~~~~~0 & ~~~~~~0 & ~~~~~0 & ~~~~~0 & ~~~~~0 \\
    * & ~~* & ~~~~* & ~~~~* & ~~~~~0 & ~~~~~~0 & ~~~~~0 & ~~~~~0 & ~~~~~0 \\
    * & ~~* & ~~~~* & ~~~~* & ~~~~~0 & ~~~~~~0 & ~~~~~0 & ~~~~~0 & ~~~~~0 \\
    0& ~~0 & ~~~~0 & ~~~~0 & ~~~~~* & ~~~~~~0 & ~~~~~0 & ~~~~~0 & ~~~~~0 \\
    0& ~~0 & ~~~~0 & ~~~~0 & ~~~~~0 & ~~~~~~* & ~~~~~* & ~~~~~0 & ~~~~~0 \\
    0& ~~0 & ~~~~0 & ~~~~0 & ~~~~~0 & ~~~~~~* & ~~~~~* & ~~~~~0 & ~~~~~0 \\
    0& ~~0 & ~~~~0 & ~~~~0 & ~~~~~0 & ~~~~~~0 & ~~~~~0 & ~~~~~* & ~~~~~* \\
    0& ~~0 & ~~~~0 & ~~~~0 & ~~~~~0 & ~~~~~~0 & ~~~~~0 & ~~~~~* & ~~~~~*
    \end{bmatrix}.\nonumber \\
        &\begin{array}{@{}*{9}{wc{7mm}}@{}}
~~|11) & |22) & |12) & |21) & |33) & |13) & |23) & |31) & |32)\\ 
\label{eq:H0-matrix-2}
\end{array} 
\end{align}
Note that we have
\begin{align}
    (33|\widetilde{U}_0(s)|33)=\frac{1}{s+\lambda}.
    \label{eq:U033}
\end{align}

To proceed, let us note from Eq.~\eqref{eq:DeltaT} that $\Delta \mathcal{T}$ operates only within the $4 \times 4$ Hilbert space spanned by the states $|22)$, $|33)$, $|23)$ and $|32)$. Then, using Eq.~\eqref{eq:H0-matrix-2}, we evidently have for the second term in the series in Eq.~\eqref{eq:series-1} that
\begin{align}
 (11|\widetilde{U}_0(s)\lambda \Delta \mathcal{T}\widetilde{U}_0(s)|11)&=\lambda (11|\widetilde{U}_0(s)|22) (22|\Delta \mathcal{T}|22)(22|\widetilde{U}_0(s)|11)\nonumber \\
&=\frac{\lambda}{2}(11|\widetilde{U}_0(s)|22)(22|\widetilde{U}_0(s)|11).
\end{align}
Similarly, the third term in the series reads as
\begin{align}
 &\lambda^2(11|\widetilde{U}_0(s)\Delta \mathcal{T}\widetilde{U}_0(s)\Delta \mathcal{T}\widetilde{U}_0(s)|11)\nonumber \\
 &=\lambda^2(11|\widetilde{U}_0(s)|22)\nonumber \\
 &\times \left[(22|\Delta \mathcal{T}|22)(22|\widetilde{U}_0(s)|22)(22|\Delta \mathcal{T}|22)+(22|\Delta \mathcal{T}|33)(33|\widetilde{U}_0(s)|33)(33|\Delta \mathcal{T}|22)\right](22|\widetilde{U}_0(s)|11)\nonumber \\
 &=\left(\frac{\lambda}{2}\right)^2(11|\widetilde{U}_0(s)|22) \left[(22|\widetilde{U}_0(s)|22)+(33|\widetilde{U}_0(s)|33)\right](22|\widetilde{U}_0(s)|11).
\end{align}
The fourth term in the series reads as
\begin{align}
 &\lambda^3(11|\widetilde{U}_0(s)\Delta \mathcal{T}\widetilde{U}_0(s)\Delta \mathcal{T}\widetilde{U}_0(s)\Delta \mathcal{T}\widetilde{U}_0(s)|11)\nonumber \\
 &=\lambda^3(11|\widetilde{U}_0(s)|22)\Big[\Big((22|\Delta \mathcal{T}|22)(22|\widetilde{U}_0(s)|22)(22|\Delta \mathcal{T}|22)(22|\widetilde{U}_0(s)|22)\nonumber \\
 &+ (22|\Delta \mathcal{T}|33)(33|\widetilde{U}_0(s)|33)(33|\Delta \mathcal{T}|22)(22|\widetilde{U}_0(s)|22)\Big) (22|\Delta \mathcal{T}|22)\nonumber \\
 &+\Big((22|\Delta \mathcal{T}|22)(22|\widetilde{U}_0(s)|22)(22|\Delta \mathcal{T}|33)(33|\widetilde{U}_0(s)|33)\nonumber \\
 &+ (22|\Delta \mathcal{T}|33)(33|\widetilde{U}_0(s)|33)(33|\Delta \mathcal{T}|33)(33|\widetilde{U}_0(s)|33)\Big) (33|\Delta \mathcal{T}|22)\Big](22|\widetilde{U}_0(s)|11)\nonumber \\
 &=\left(\frac{\lambda}{2}\right)^3(11|\widetilde{U}_0(s)|22) \left[(22|\widetilde{U}_0(s)|22)+(33|\widetilde{U}_0(s)|33)\right]^2(22|\widetilde{U}_0(s)|11),
\end{align}
and so on. Summing up all the terms in the series, we finally arrive at our main result
\begin{align}
&\widetilde{P}_1(s)=(11|\widetilde{U}_0(s)|11)+\frac{\lambda}{2}(11|\widetilde{U}_0(s)|22)(22|\widetilde{U}_0(s)|11)\left[1-\frac{\lambda}{2}\left((22|\widetilde{U}_0(s)|22)+(33|\widetilde{U}_0(s)|33)\right)\right]^{-1}\nonumber \\
&=(11|\widetilde{U}_0(s)|11)+\frac{\lambda}{2}(11|\widetilde{U}_0(s)|22)(22|\widetilde{U}_0(s)|11)\left[1-\frac{\lambda}{2}\left((22|\widetilde{U}_0(s)|22)+\frac{1}{s+\lambda}\right)\right]^{-1},
\label{eq:P1s}
\end{align}
where in obtaining the last equality, we have used Eq.~\eqref{eq:U033}.

\subsubsection{Matrix elements of superoperator $\widetilde{U}_0(s)$}
\label{sec:U0s}

In order to proceed, we need to first compute the quantity $(11|\widetilde{U}_0(s)|11)$. Let us start with Eq.~\eqref{eq:Us-3} and write it as 
\begin{align}
\widetilde{U}_0(s)=\mathcal{G}+\lambda \mathcal{G} \mathcal{T}_1\mathcal{G}+\lambda^2 \mathcal{G} \mathcal{T}_1\mathcal{G}\mathcal{T}_1\mathcal{G}+\lambda^3 \mathcal{G} \mathcal{T}_1\mathcal{G}\mathcal{T}_1\mathcal{G}\mathcal{T}_1\mathcal{G}+\ldots, 
\label{eq:calc-1}
\end{align}
where we have defined
\begin{align}
 \mathcal{G} \equiv \left[(s+\lambda)\mathbb{I}+i H_0^\times\right]^{-1}.
        \label{eq:Us-4}
\end{align}

Let us rewrite the Hamiltonian $H_0$ in Eq.~\eqref{eq:H0-matrix-1} thus:  defining $\epsilon \equiv (\epsilon_1+\epsilon_2)/2$ as the average energy of levels $1$ and $2$ and $\delta \epsilon \equiv (\epsilon_1-\epsilon_2)/2$ as one-half of the energy gap between levels $1$ and $2$, we get on neglecting a constant term given by $\epsilon \mathbb{I}$ that
\begin{align}
    H_0=\begin{bmatrix} \delta \epsilon& \Delta & 0 \\ \Delta & -\delta \epsilon & 0 \\ 0 &0& 0 \end{bmatrix},
    \label{eq:H0-matrix}
\end{align}
so that we have 
\begin{align}
    H_0^\times=\begin{array}{c@{}}
(11| \\ (22| \\ (12| \\ (21| \\ (33| \\ (13| \\ (23| \\ (31| \\ (32|
\end{array}
&\begin{bmatrix} 
    0& 0 & -\Delta & \Delta & ~~~0 & ~~~~0 & ~~~0 & ~~~0 & ~~~0 \\
    0& 0 & \Delta & -\Delta & ~~~0 & ~~~~0 & ~~~0 & ~~~0 & ~~~0 \\
    -\Delta & \Delta & 2\delta \epsilon & 0 & ~~~0 & ~~~~0 & ~~~0 & ~~~0 & ~~~0 \\
    \Delta & -\Delta & 0 & -2\delta \epsilon & ~~~0 & ~~~~0 & ~~~0 & ~~~0 & ~~~0 \\
    0& 0 & 0 & 0 & ~~~0 & ~~~~0 & ~~~0 & ~~~0 & ~~~0 \\
    0& 0 & 0 & 0 & ~~~0 & ~~~~\delta\epsilon & ~~~\Delta & ~~~0 & ~~~0 \\
    0& 0 & 0 & 0 & ~~~0 & ~~~~\Delta & ~~~-\delta \epsilon & ~~~0 & ~~~0 \\
    0& 0 & 0 & 0 & ~~~0 & ~~~~0 & ~~~0 & ~~~-\delta \epsilon & ~~~-\Delta \\
    0& 0 & 0 & 0 & ~~~0 & ~~~~0 & ~~~0 & ~~~-\Delta & ~~~\delta \epsilon
    \end{bmatrix}.\nonumber \\
        &\begin{array}{@{}*{9}{wc{7mm}}@{}}
~~~~~~|11) & ~~~~~~|22) & ~~~~~~|12) & ~~~~~~|21) & ~~~~~~~~~~|33) & ~~~~~~~~~~~~|13) & ~~~~~~~~~~~~~~~~|23) & ~~~~~~~~~~~~~~~~~~~~~~~~~~|31) & ~~~~~~~~~~~~~~~~~~~~~~~~~~~~~~~~|32)\\ 
\label{eq:H0-matrix-2-2}
\end{array} 
\end{align}

On the other hand, for later convenience, we write the operator $S$ in Eq.~\eqref{eq:Vp} as  
\begin{align}
    S=\begin{bmatrix} 0 & \mathbf{0}^\mathrm{T} \\ \mathbf{0} & \sigma_x   \end{bmatrix},
\end{align}
with $\mathbf{0}\equiv \begin{bmatrix} 0 \\ 0 \end{bmatrix}$, $\mathrm{T}$ denoting matrix transpose operation, and $\sigma_x$ the usual Pauli matrix. 

We now come back to evaluation of the various terms in Eq.~\eqref{eq:calc-1}. Noting that $\mathcal{T}_1$ has its only nonzero element given by $(11|\mathcal{T}_1|11)=1$, as discussed earlier, we get 
\begin{align}
(nm|\lambda \mathcal{G} \mathcal{T}_1\mathcal{G}|n'm')&=\lambda\sum_{n'',m'',n''',m'''} (nm|\mathcal{G}|n''m'')(n''m''|  \mathcal{T}_1|n'''m''')(n'''m'''|\mathcal{G}|n'm')\nonumber \\
&=\lambda (nm|\mathcal{G}|11)(11|\mathcal{G}|n'm').
\end{align}
A similar exercise may be carried out for the other terms in Eq.~\eqref{eq:calc-1}, leading to the result
\begin{align}
&(nm|\widetilde{U}_0(s)|n'm')&\nonumber \\
&=(nm|\mathcal{G}|n'm')+\lambda (nm|\mathcal{G}|11)(11|\mathcal{G}|n'm')+\lambda^2 (nm|\mathcal{G}|11)(11|\mathcal{G}|11)(11|\mathcal{G}|n'm')\nonumber \\
&+\lambda^3 (nm|\mathcal{G}|11)(11|\mathcal{G}|11)(11|\mathcal{G}|11)(11|\mathcal{G}|n'm')+\ldots\nonumber \\
&=(nm|\mathcal{G}|n'm')+\lambda (nm|\mathcal{G}|11)[1-\lambda (11|\mathcal{G}|11)]^{-1}(11|\mathcal{G}|n'm'), 
\label{eq:calc-2}
\end{align}
so that we have 
\begin{align}
&(11|\widetilde{U}_0(s)|11)=(11|\mathcal{G}|11)[1-\lambda (11|\mathcal{G}|11)]^{-1},\nonumber \\
&(11|\widetilde{U}_0(s)|22)=(11|\mathcal{G}|22)[1-\lambda (11|\mathcal{G}|11)]^{-1},\nonumber \\
&(22|\widetilde{U}_0(s)|11)=(22|\mathcal{G}|11)[1-\lambda (11|\mathcal{G}|11)]^{-1},\nonumber \\
&(22|\widetilde{U}_0(s)|22)=(22|\mathcal{G}|22)+\lambda (22|\mathcal{G}|11)(11|\mathcal{G}|22)[1-\lambda (11|\mathcal{G}|11)]^{-1}.\nonumber \\
\label{eq:G-nonzero}
\end{align}

To proceed, we must compute the quantity $(nm|\mathcal{G}|n'm')$, which is done easily by reverting to the time domain:
\begin{align}
 (nm|\mathcal{G}|n'm')&=\int_0^\infty dt~e^{-(s+\lambda)t}(nm|e^{-iH_0^\times t}|n'm')\nonumber \\
 &=\int_0^\infty dt~e^{-(s+\lambda)t}\langle n|e^{-iH_0t}|n'\rangle \langle m'|e^{iH_0 t}|m\rangle.   
\end{align}
Note that the $2\times 2$ block with non-zero entries in the expression for $H_0$ in Eq.~\eqref{eq:H0-matrix} may be written as equal to $\delta \epsilon \sigma_z+\Delta \sigma_x$. Moreover, using the following result that follows straightforwardly from the properties of the Pauli matrices,
\begin{align}
    e^{\theta\mathrm{i}\mathbf{v}\cdot \mathbf{\sigma}}=\cos(\theta \alpha)\mathbb{I}+\mathrm{i}\sin(\theta \alpha) \frac{\mathbf{v}\cdot \mathbf{\sigma}}{\sqrt{\alpha}};~\alpha\equiv \sqrt{\sum_{i=1}^3 v_i^2},
\end{align}
we get
\begin{align}
e^{-iH_0 t}=\cos(\omega t)\mathbb{I}-\mathrm{i}\sin (\omega t) \frac{\delta \epsilon \sigma_z + \Delta \sigma_x}{\omega};~\omega \equiv \sqrt{(\delta \epsilon)^2+\Delta^2}. 
\end{align} 
This yields \begin{align}
    (11|\mathcal{G}|11)=(22|\mathcal{G}|22)&=\int_0^\infty dt~e^{-(s+\lambda)t}[\cos^2(\omega t)+(\delta \epsilon/\omega)^2\sin^2(\omega t)]\nonumber \\
    &=\frac{(s+\lambda)^2+2[\omega^2+(\delta \epsilon)^2]}{(s+\lambda)[(s+\lambda)^2+4\omega^2]},\nonumber \\\label{eq:U0s-nonzero} \\
    (11|\mathcal{G}|22)=(22|\mathcal{G}|11)&=\int_0^\infty dt~e^{-(s+\lambda)t}(\Delta/\omega)^2\sin^2(\omega t)\nonumber \\
    &=\frac{2\Delta^2}{(s+\lambda)[(s+\lambda)^2+4\omega^2]}.\nonumber
\end{align}

We therefore arrive at the following expressions:
\begin{align}
&(11|\widetilde{U}_0(s)|11)=(11|\mathcal{G}|11)[1-\lambda (11|\mathcal{G}|11)]^{-1},\\
&(22|\widetilde{U}_0(s)|11)=(11|\widetilde{U}_0(s)|22)=\frac{2\Delta^2}{(s+\lambda)[(s+\lambda)^2+4\Delta^2+4(\delta \epsilon)^2]}[1-\lambda (11|\mathcal{G}|11)]^{-1},\\
&(22|\widetilde{U}_0(s)|22)=\frac{(s+\lambda)^2+2[\Delta^2+2(\delta \epsilon)^2]}{(s+\lambda)[(s+\lambda)^2+4\Delta^2+4(\delta \epsilon)^2]}\nonumber \\
&\hspace{2.75cm}+\lambda \left(\frac{2\Delta^2}{(s+\lambda)[(s+\lambda)^2+4\Delta^2+4(\delta \epsilon)^2]}\right)^2[1-\lambda (11|\mathcal{G}|11)]^{-1}, \\
&(11|\mathcal{G}|11)=\frac{(s+\lambda)^2+2[\Delta^2+2(\delta \epsilon)^2]}{(s+\lambda)[(s+\lambda)^2+4\Delta^2+4(\delta \epsilon)^2]},
\end{align}
which when used in Eq.~\eqref{eq:P1s} followed by performing an inverse Laplace transform yields the desired stay-put probability $P_1(t)$.

\subsubsection{Checking our results on stay-put probability for a few limiting cases}
\label{sec:check}

\begin{enumerate}
\item Let us check our results for the case $\lambda=0$, when the system has only unitary evolution. From our analysis, we get
\begin{align}
    \widetilde{P}_1(s)=(11|\mathcal{G}|11)=\frac{s^2+2[\Delta^2+2(\delta \epsilon)^2]}{s[s^2+4\Delta^2+4(\delta \epsilon)^2]},
\end{align}
 on using the first equation in Eq.~\eqref{eq:U0s-nonzero}. We then get
\begin{align}
    P_1(t)=\frac{\Delta^2\cos^2 \left(t\sqrt{\Delta^2+(\delta \epsilon)^2}\right)+(\delta \epsilon)^2}{\Delta^2+(\delta \epsilon)^2} \le 1.
    \label{eq:Rabi-result}
\end{align}
With $\lambda=0$, the level $3$ remains inaccessible, and hence we have $P_1(t)+P_2(t)=1$, yielding
\begin{align}
    P_2(t)=\frac{\Delta^2\sin^2 \left(t\sqrt{\Delta^2+(\delta \epsilon)^2}\right)}{\Delta^2+(\delta \epsilon)^2} \le 1.
\end{align}
The above are the expressions for the celebrated Rabi oscillations, describing the time development of the population of a two-level system undergoing unitary evolution, see Eq.~\eqref{eq:Rabi}.

\item The next limit we want to consider is $\Delta=0$. In this case, it is easily checked that we have $(22|\widetilde{U}_0(s)|11)=(11|\widetilde{U}_0(s)|22)=0$, and that $(22|\widetilde{U}_0(s)|22)=(11|\mathcal{G}|11)=1/(s+\lambda)$, yielding 
$(11|\widetilde{U}_0(s)|11)=1/s$. We thus get $\widetilde{P}_1(s)=(11|\widetilde{U}_0(s)|11)=1/s$, resulting in $P_1(t)=1$. This is expected; indeed, the system is initially in level $1$, but the absence of the $\Delta$ term does not induce any transition to level $2$, and hence, application of pulses inducing transitions between levels $2$ and $3$ is essentially ineffective in bringing about any change in the state of the system. Consequently, one has $P_1(t)=1~\forall~t$. 
\end{enumerate}

\subsubsection{Numerical results on stay-put probability}
\label{sec:numerical-results}

\begin{figure}
   \includegraphics[width=0.7\linewidth,left]{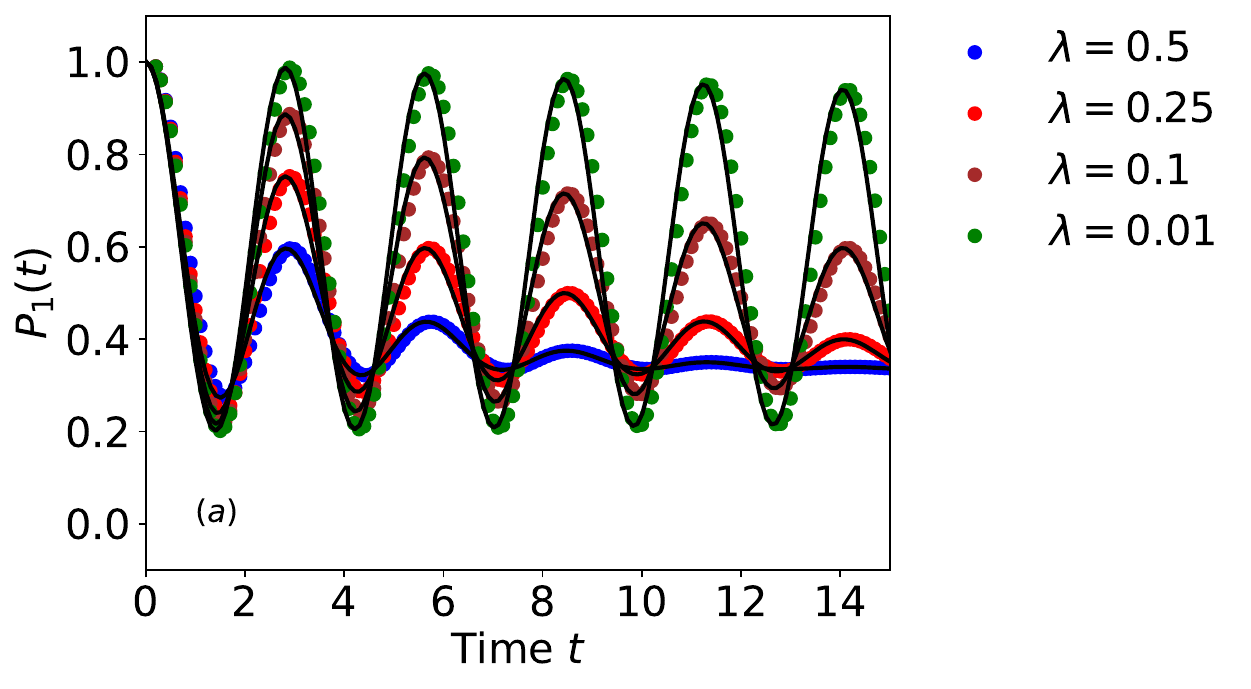}\\
   \includegraphics[width=0.5\linewidth,left]{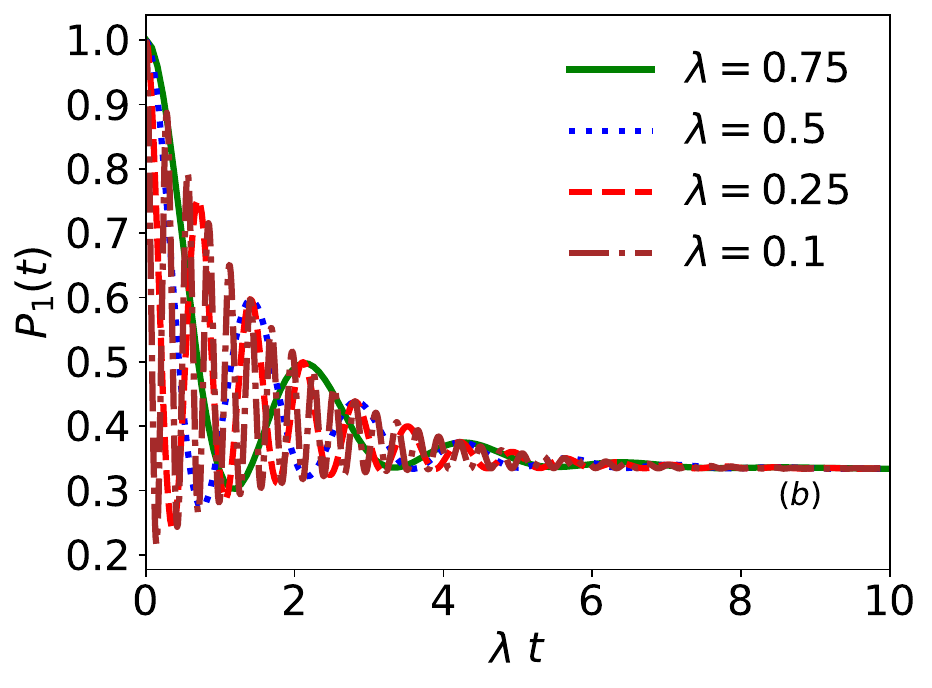}\\
    \includegraphics[width=0.5\linewidth,left]{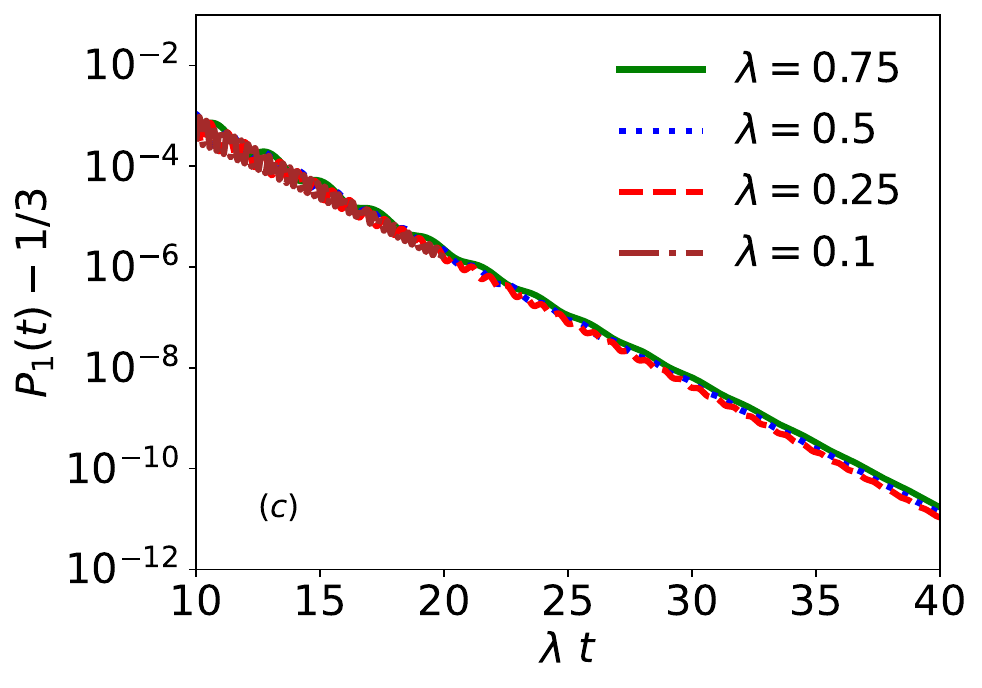}
    \caption{(a) Stay-put probability $P_1(t)$ as a function of time $t$ for different $\lambda$'s, with $\Delta=1.0$ and $\delta \epsilon=0.5$. Here, lines correspond to data obtained by evaluating numerically Eq.~\eqref{eq:rhot-evaluate}, while the data for the points are obtained by evaluating numerically the inverse Laplace transform of our exact analytical result, Eq.~\eqref{eq:P1s}. (b) The data plotted as a function of $\lambda t$; the scaling collapse of the data for different $\lambda$ and at long times, together with the data in panel (c), suggests the scaling $P_1(t)-1/3 \sim e^{-\lambda t}$, for large $t$.}
    \label{fig:fig1}
\end{figure}

\begin{figure}
   \includegraphics[width=0.5\linewidth]{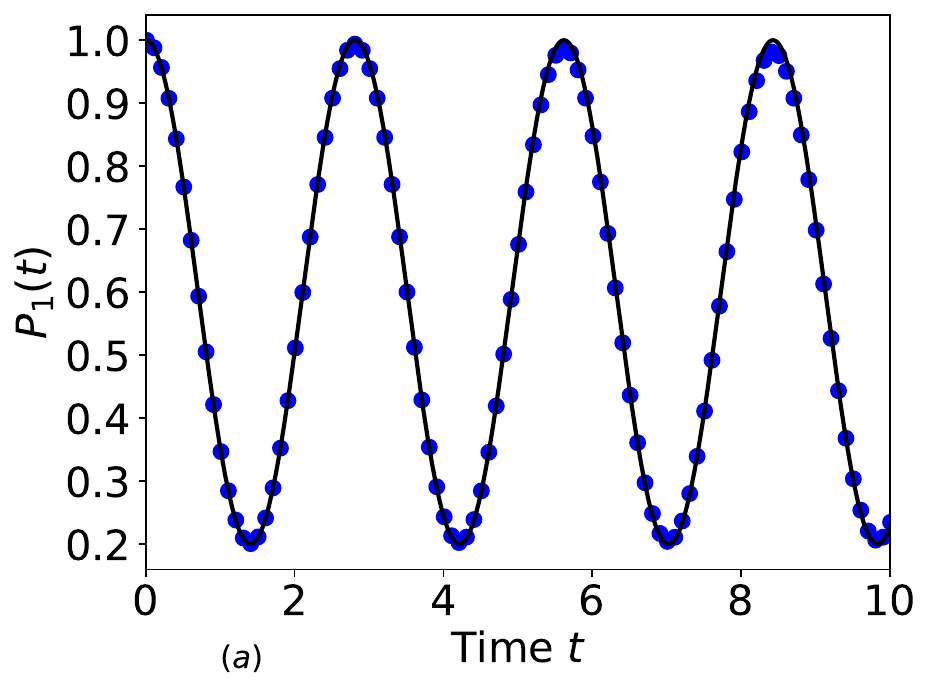}
   \includegraphics[width=0.5\linewidth]{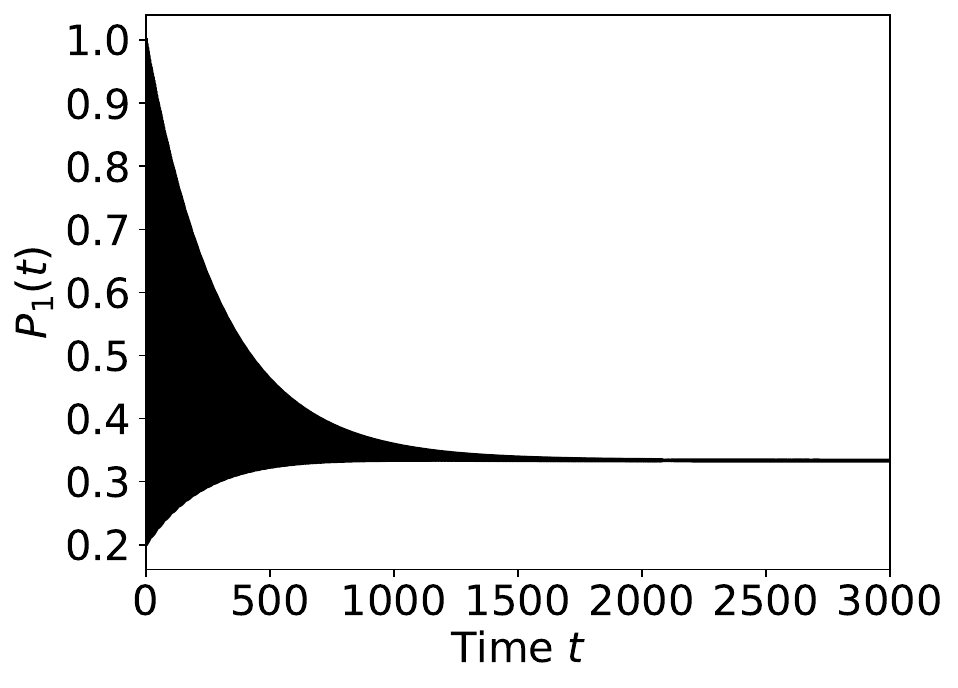}
    \caption{Stay-put probability $P_1(t)$ as a function of time $t$ at small times $t\ll 1/\lambda$ (panel (a)) and at long times $t\gg 1/\lambda$ (panel (b)), with $\Delta=1.0$, $\delta \epsilon=0.5$, $\lambda=0.005$. The data (points in panel (a) and line in panel (b)) are obtained by evaluating numerically the inverse Laplace transform of our exact analytical result, Eq.~\eqref{eq:P1s}. The line in panel (a) corresponds to Eq.~\eqref{eq:Rabi-result}. The plots demonstrate a crossover between a low-$\lambda$ coherence regime and a high-$\lambda$ decoherence regime.}
    \label{fig:fig2}
\end{figure}

Here, we provide explicit results on stay-put probability, based on numerical Laplace inversion of our exact analytical result, Eq.~\eqref{eq:P1s}. The first thing we do is to check the validity of our analytical results, by comparing them with those obtained by evolving numerically the initial density operator $\rho(0)$ in Eq.~\eqref{eq:IC} according to Eq.~\eqref{eq:rhot-evaluate}. In Fig.~\ref{fig:fig1}(a), we demonstrate a match between the two results, thereby validating our analytical results. In panel (b), we plot based on our analytical results the stay-put probability as a function of $\lambda t$ for given values of $\Delta$ and $\delta \epsilon$ and for different values of $\lambda$. The scaling collapse of the data at large times implies the scaling $P_1(t)=f(\lambda t)$ at large times. In panel (c), we show that in fact, one has $P_1(t)-1/3 \sim \exp(-\lambda t)$ for large $t$. That $P_1(t)$ decays at long times to the value $1/3$ is consistent with Eq.~\eqref{eq:rhoasymp}. On the basis of results presented in Fig.~\ref{fig:fig2}, we conclude that while for $t\ll 1/\lambda$, we have $P_1(t)$ given by Eq.~\eqref{eq:Rabi-result}, a result that is an outcome of the coherent evolution of the system, for $t\gg 1/\lambda$, one has instead that $P_1(t)-1/3\sim \exp(-\lambda t)$, which stems from decoherence induced into the dynamics as a result of application of pulses at random times and of random strengths.  We thus have a clear demonstration of coherence-to-decoherence crossover due to coupling with the external environment, a phenomenon that we demonstrate for our model system on the basis of exact analytical results derived using the stochastic Liouville equation approach.  

\section{Conclusion}
\label{sec:conclusion}

In the present investigation, we have focused on the effective time-evolution operator in the Schr\"{o}dinger picture of the density operator for a generic three-level quantum system. The formulation is based on how a typical quantum measurement is performed in which an otherwise unitary evolution under a Hermitian Hamiltonian is subject to (random) influences of a measuring apparatus~\cite{Gherardini:2017}. Although the underlying method is reminiscent of the well-established concepts of Rabi oscillations, magnetic resonance and stochastic theory of spectral line shapes, it has ramifications for more modern quantum trajectory schemes, applied in mostly quantum optics literature~\cite{Wiseman:2011}. In that respect, the presently undertaken three-level scenario can be contrasted with yet another distinct attempt in which all the three levels belong to the coherently-coupled subsystem, wherein levels $2$ and $3$ are nearly degenerate but differently populated with respect to level $1$ because of finite-temperature effects~\cite{Das:2025}. The latter attempt hinges on how degeneracy slows down decoherence effects, a desired objective for retaining quantum information. Furthermore, the current treatment has an overlap with our earlier analyses of projective measurements~\cite{Das:2022} and quantum resetting~\cite{Dattagupta:2022} in the context of the tight-binding chain dynamics that models the motion of a quantum particle between the sites of a one-dimensional lattice. The former refers to projective measurements performed at random times by means of a detector that detects the presence of the quantum particle on a target site, while the latter comprises a dynamics in which the density operator is at random times reset to its initial form~\cite{Mukherjee:2018}. It is worthwhile to mention that both the aforementioned set-ups are distinctly different from the one studied in the current work.

We now comment on some relevant future directions of the current work. The Rabi oscillations discussed in the Introduction, and whose manifestation is seen in the behavior of the quantity $P_1(t)$ for our studied model, can also be discerned in a generalized version of Eq.~\eqref{eq:Rabi-1} in which the field in the $xy$-plane is an oscillatory one that couples to both  $x$- and $y$-components of the transverse field~\cite{Dattagupta1}. In this case, Eq.~\eqref{eq:Rabi-1} acquires a parametric time-dependence:
\begin{align}
H_0(t)=-B_0\sigma_z - B (\sigma_x \cos (\omega t) + \sigma_y \sin (\omega t)). 
\label{eq:Rabi-2}
\end{align}
The consequent phenomena are relevant for magnetic resonance set-up, either of the nuclear type (NMR) or of the electron type (ESR)~\cite{Dattagupta:1987}. By going to a frame that rotates around the $z$-axis, the underlying Hamiltonian can be brought to the time-independent structure of Eq.~\eqref{eq:Rabi-1}, and we can analytically carry out an identical treatment of the underlying Rabi oscillations~\cite{Sakurai:1965}. It would be an interesting poser to address the question of what transpires when a pulse sequence of the kind studied in this work (with the average frequency of application of pulses being $\lambda$) is introduced in the rotating frame, which would naturally connect the problem to considerations of relaxation phenomena in magnetic resonance. While in the magnetic resonance context, the frequency $\omega$ in Eq.~\eqref{eq:Rabi-2} must be matched with the Zeeman frequency associated with the static field $B$ in order to achieve the resonance condition, we would however be interested in a more general situation in which the frequency $\omega$ can be arbitrary. We would then, in principle, be able to map the entire domain of the interplay of $\omega$ and $\lambda$. If there is no decoherence (i.e., $\lambda = 0$), there is the intriguing occurrence of a Berry phase, and it would be interesting to explore how presence of $\lambda$ disrupts this additional quantum phase~\cite{Dattagupta1}. Furthermore, there is also the emergence of a separate and fascinating Kapitza phenomenon when $\omega$ is arbitrarily large and the underlying spin feels an effective static field~\cite{Dattagupta2}. We would like to examine how the pulse sequencing tampers with the Kapitza limit. However, the frequency-dependent case is a separate one and is relegated to a future study.

\section{Acknowledgments}We are very grateful to the anonymous referee for pointing out an error in our computation. S.D. thanks the Indian National Science Academy for support
through their Honorary Scientist Scheme.
S.G. thanks ICTP–Abdus Salam International Centre for Theoretical Physics, Trieste, Italy, for support under its Regular Associateship scheme, and for hospitality in March 2025 when the paper was finalized. S.G. also thanks Anish Acharya, Rupak Majumder, and Soumya Kanti Pal for help with numerical computation and preparation of the figures. We gratefully acknowledge generous allocation of computing resources by the Department of Theoretical Physics (DTP) of the Tata Institute of Fundamental Research (TIFR), and related technical assistance from Kapil Ghadiali and Ajay Salve. This work is supported by the Department of Atomic Energy, Government of India, under Project Identification Number RTI 4002.

\end{document}